# A phase transition driven by subtle distortion without broken symmetry on spin, charge and lattice in Layered $Ln$Cu$_{4-\delta}$P$_2$ ($Ln$=Eu, Sr)

Yong Nie[1,2,#], Zheng Chen[1,#], Wensen Wei[1], Huijie Li[1], Yong Zhang[2], Ming Mei[1,2], Wenhai Song[3], Dongsheng Song[4], Wei Ning[1,*], Zhaosheng Wang[1,*], Xiangde Zhu[1,*], Mingliang Tian[1,5]

[1]*Anhui Key Laboratory of Condensed Matter Physics at Extreme Conditions, High Magnetic Field Laboratory, HFIPS, Anhui, Chinese Academy of Sciences, Hefei 230031, China*

[2]*Department of Physics, University of Science and Technology of China, Hefei 230031, Anhui, China*

[3]*Key Laboratory of Materials Physics, Institute of Solid State Physics, HFIPS, Chinese Academy of Sciences, Hefei 230031, China*

[4] *Information Materials and Intelligent Sensing Laboratory of Anhui Province, Key Laboratory of Structure and Functional Regulation of Hybrid Materials of Ministry of Education, Institutes of Physical Science and Information Technology, Anhui University, Hefei 230601, China*

[5]*School of Physics, Anhui University, Hefei 230601, China*

In the scenario of Landau phase transition theory in condensed matter physics, any thermal dynamic phase transition must be subject to some kind of broken symmetries, that are relative to its spin, charge, orbital and lattice. Here we report a rare phase transition at $T_p$ ~120 K or 140 K in layered materials $Ln$Cu$_{4-\delta}$P$_2$ ($Ln$=Eu, Sr) driven by a subtle structural-distortion without any broken symmetry on charge, spin and lattice. The variations of the lattice parameters, ($\Delta L_c/L_c$) ~ 0.013% or 0.062%, verified by thermal expansion, is much less than that for a typical crystalline phase transition (~0.5-1%), but the significant anomaly in heat capacity provides clear evidence of its intrinsic nature of thermodynamic transition.

Key words: Symmetry Breaking; Transition; Correlated Electronic System;

#: These authors contribute equally.

*:Corresponding authors' email:

ningwei@hmfl.ac.cn; zswang@hmfl.ac.cn; xdzhu@hmfl.ac.cn

In strongly correlated electron systems, there exists very rich fascinating phenomena due to the coupling of the order parameters among orbital, charge, lattice and spin [1,2], such as anti-ferromagnetism (AFM) [3,4], superconductivity (SC) [5–7], charge or orbital ordering (CO)/charge density wave (CDW) [8–14], and colossal magnetoresistance (CMR) [15–20] and so on. By chemical doping or applying pressure, these orders can be tuned, showing coexisting and/or competing phase diagrams. Upon chemical substitution in cuprate oxides, the AFM ground state can be suppressed, accompanied by appearance of SC and electronic stripe phases at appropriate doping region [21]. For superconducting iron arsenide $BaFe_{2-x}Co_xAs_2$, in addition to the AFM ground state for parent compound, a small structural distortion breaks the 'C4' rotational symmetry in the underdoped region of the phase diagram [22]. Charge-orbital density wave induced phase transition with broken translational symmetry was observed in $IrTe_2$ [23,24] while an orbital order transition with broken rotational symmetry was proved in $MgV_2O_4$ [25]. Interestingly, CrAs endures an anti-ferromagnetic transition with both broken rotational/

translational symmetry accompanying by a significant volume change [26,27]. Recently, a layered magnetic material EuAg$_4$As$_2$ has been reported with a first-order structural transition at ~ 120 K, followed by a superstructure distortion verified by the single-crystal X-ray precession images below 120 K [28].

$Ln$Cu$_{4-\delta}$P$_2$ ($Ln$ =Eu and Sr) is iso-structural to EuAg$_4$As$_2$, and adopts the space group of $R\bar{3}m$ in rhombohedral CaCu$_4$P$_2$ structure [29]. The crystal structure can be regarded as inserting an additional itinerant Cu$_2$ layer within the Cu$_2$P$_2$ layer of the trigonal CaAl$_2$Si$_2$-type stacking of -$Ln$-Cu$_2$P$_2$-$Ln$-. Cu ion is partially occupied in $Ln$Cu$_{4-\delta}$P$_2$ as the same as Ag ion in EuAg$_4$As$_2$ [28]. In this report, we performed detailed structural characterization and physical property measurements on $Ln$Cu$_{4-\delta}$P$_2$ single crystals by transmission electron microscope (TEM), electrical transport, magnetization, specific heat and thermal expansion. We revealed an intriguing transition at 120 K in EuCu$_{4-\delta}$P$_2$ and at 140 K in SrCu$_{4-\delta}$P$_2$, where both resistivity and heat capacity show significant anomaly, but no any signature of broken symmetry on its charge order or translational/rotational symmetry were detected. This result is in sharp contrast to that in EuAg$_4$As$_2$ reported previously. We attribute this transition probably in relation to a subtle change of orbital order from Cu-P layer.

Single crystals $Ln$Cu$_{4-\delta}$P$_2$ were grown via Bi-Cu flux method. The stoichiometry of starting raw materials of Eu pieces (3N purity)/ (Sr shots (3N purity)), Cu plates (4N purity), red phosphorus pieces (3N purity) and Bi shots (5N purity) is 1:8:2:35 in molar ratio. Oxide layer of Eu pieces/Sr shots were scraped off from the surface. The mixture was placed inside an alumina crucible and sealed in a quartz tube, slowly heated to 1050 ℃, then kept at this temperature for 20 h, and finally cooled to 850 ℃ at cooling rate of 3 ℃/h before separating the flux in a centrifuge. Phase purity was checked by X-ray diffraction using Rigaku-TTR3 with Cu-$K_\alpha$ radiation. The chemical stoichiometry for both samples are determined by energy dispersive spectrum (EDS) equipped upon Helios nanolab600i. Resistivity and specific heat measurements were performed upon a Physical Properties Measurement System (PPMS-14T, QD Inc.). Magnetization measurement was performed upon a Magnetic Properties Measurement System (MPMS3-7T, QD Inc.). The TEM and selected area electrons diffraction (SAED) were performed by Talos F200X equipped with a liquid-nitrogen-cooled holder. Thermal expansions were measured by home-made strain gauge [30] and capacitance dilatometer [31], respectively.

The layered compound $Ln$Cu$_{4-\delta}$P$_2$ crystallizes in rhombohedral CaCu$_4$P$_2$ structure, as illustrated in Fig. 1(a) [29]. Eu ions (Sr ions) are located between Cu$_2$P layers. Fig. 1(b) and 1(c) shows the XRD patterns of the single crystals of EuCu$_{4-\delta}$P$_2$ and SrCu$_{4-\delta}$P$_2$ respectively. The diffraction peaks of the crystal face are well indexed with (00$l$) plane, which is the same with those reported in EuAg$_4$As$_2$ [28,32]. As shown in the inset of Fig. 1(b) and 1 (c), the full width at half maximum (FWHM) of the rocking curve of (006)/(003) peaks reaches as low as 0.08/0.24 degree, indicating the high quality of the as-grown single crystals.

Firstly, we have measured the magnetic properties of EuCu$_{4-\delta}$P$_2$ and SrCu$_{4-\delta}$P$_2$. Fig. 1(d) shows the zero-field-cooling (ZFC) and field-cooling (FC) temperature-dependent magnetization of single crystal EuCu$_{4-\delta}$P$_2$ at H=100 Oe aligned in the *ab* plane (H//*ab*) and along the *c*-axis, respectively. (More detailed magnetization results are shown in Fig. S1 in the supplementary materials). A sharp increase can be identified as the AFM transition at $T_N$ =33 K. As a comparison, such a magnetic transition is absent in the temperature-dependent magnetization curves of SrCu$_{4-\delta}$P$_2$ at low temperatures, as shown in Fig. 1(e).

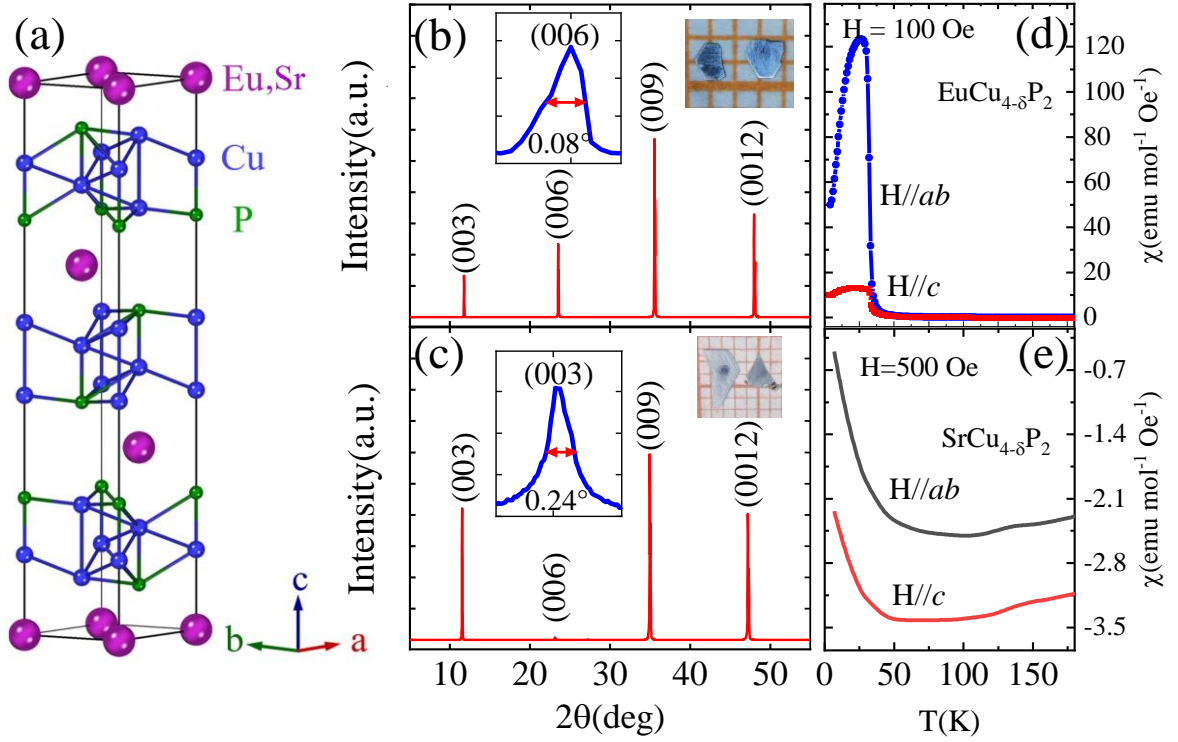

Figure 1. (a) Illustration of crystal structure of $Ln$Cu$_{4-\delta}$P$_2$ in one unit cell. (b) and (c) are the XRD patterns of the as-grown single crystals of EuCu$_{4-\delta}$P$_2$ and SrCu$_{4-\delta}$P$_2$, respectively. The left insets are the rocking curve at (006) peak of EuCu$_{4-\delta}$P$_2$ and (003) peak of SrCu$_{4-\delta}$P$_2$ respectively. The right insets are optical images of several typical single crystals. (d) and (e) are the temperature-dependent magnetization of EuCu$_{4-\delta}$P$_2$ and SrCu$_{4-\delta}$P$_2$, respectively.

Fig. 2(a) shows the temperature-dependent resistivity measured in both cooling and heating cycles for EuCu$_{4-\delta}$P$_2$. The $\rho_{xx}$-T curves present a metallic behavior with a peak at $T_N \sim$ 33 K and a kink at $T_p \sim$ 120 K with a small hysteresis. The peak near $T_N$ is consistent with the magnetization measurement due to the AFM transition, while the kink at $T_p$ corresponds to the bifurcation point of the 1/$\chi$-T curves for H//$ab$ and H//$c$ as shown in Fig. 2(c). In order to present the bifurcation clearly, the difference between the reciprocals of susceptibility for $H$//$c$ and $H$//$ab$ ($\Delta 1/\chi = 1/\chi_{\|c} - 1/\chi_{\|ab}$) is shown in Fig. 2(c). Blue short dash line serves as a reference line, $\Delta 1/\chi$ is only observable above $T_p$. The anisotropic effective susceptibilities for EuCu$_{4-\delta}$P$_2$ should originate from the crystal fields change at $T_p$ [33,34]. Fig. 2(b) shows the temperature-dependent resistivity measured in both cooling and heating cycles for SrCu$_{4-\delta}$P$_2$, where the $\rho_{xx}$-T curves present metallic behavior with a kink at $T_p \sim 140$ K accompanied by a clear hysteresis, but no the second anomaly at low temperatures. Fig. 2(d) shows the temperature-dependent magnetic susceptibilities $\chi \equiv M/H$ for $H//c$ and $H//ab$, a kink superimposed on the paramagnetic background is observed near $T_p$. It is clearly that the kink at $T_p$ has no relation to the magnetic moment of Eu ions by comparing the results of electric transport and magnetization in EuCu$_{4-\delta}$P$_2$ and SrCu$_{4-\delta}$P$_2$.

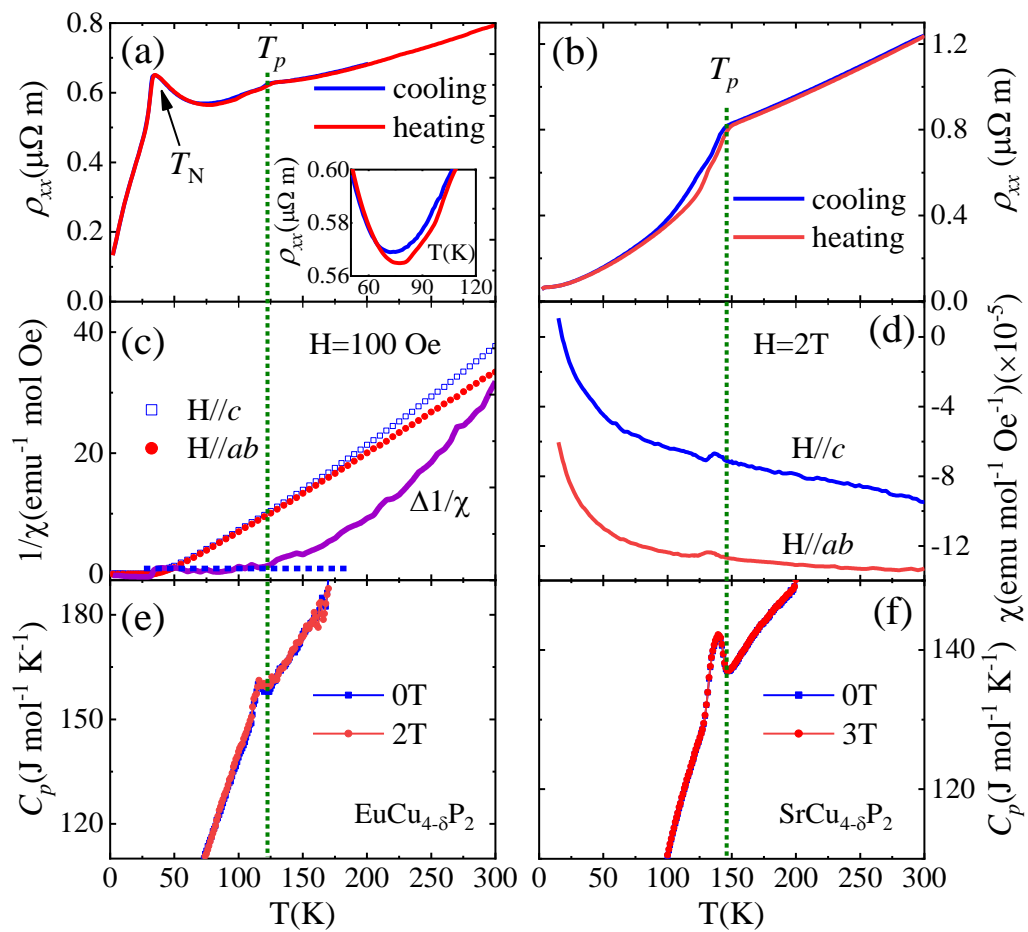

Figure 2. (a) and (b) show the temperature-dependent resistivity under cooling (blue line) and heating (red line) processes for EuCu$_{4-\delta}$P$_2$ and SrCu$_{4-\delta}$P$_2$ respectively. The inset of (a) shows the enlargement of resistivity around 90 K. (c) The temperature-dependent reciprocal of susceptibility (1/χ-T) measurement for $H//c$ and $H//ab$. The difference of the reciprocals of susceptibility for $H//c$ and $H//ab$ ($\Delta 1/\chi$) of EuCu$_{4-\delta}$P$_2$ at 100 Oe is also shown. (d) The temperature-dependent susceptibility (χ-T) curves for $H//c$ and $H//ab$ of SrCu$_{4-\delta}$P$_2$ at 2 T.

(e) and (f) are the temperature-dependent specific heat capacity ($C_p$) curves at $H = 0$ T, 2 T and $H = 0$ T, 3 T near $T_p$ in EuCu$_{4-\delta}$P$_2$ and SrCu$_{4-\delta}$P$_2$, respectively.

To figure out the nature of the anomaly in resistivity at $T_p \sim 120$ K in EuCu$_{4-\delta}$P$_2$ and 140 K in SrCu$_{4-\delta}$P$_2$, we performed temperature-dependent heat capacity ($C_p$) measurements at different magnetic fields near $T_p$ for two compounds, as shown in Fig. 2(e) and 2(f). A remarkable hump anomaly in both heat capacity was clearly revealed around $T_p$, and it shows independence on the applied magnetic field. These results unambiguously demonstrate that the anomaly near $T_p$ originates from a thermodynamic phase transition with a non-magnetic nature.

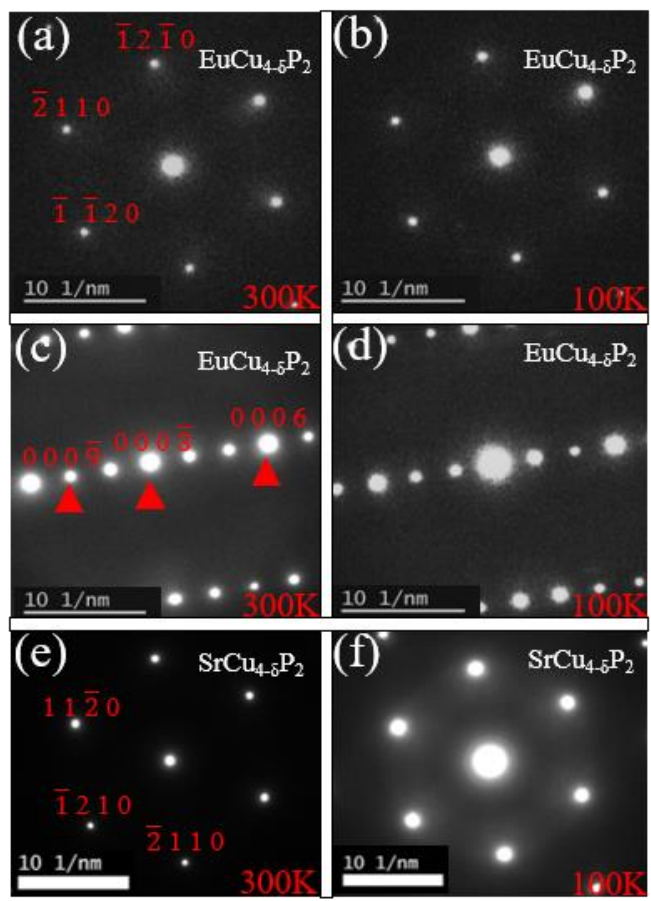

Figure 3. (a) and (b) are the SAED patterns of EuCu$_{4-\delta}$P$_2$ taken along zone axis [0001] at 300 K and 100 K, respectively. (c) and (d) are the SAED patterns of EuCu$_{4-\delta}$P$_2$ taken along $c$-axis along zone axis [10$\bar{1}$0] at 300 K and 100 K, respectively. (e) and (f) are the SAED patterns of SrCu$_{4-\delta}$P$_2$ taken along zone axis [0001] at 300 K and 100 K, respectively.

To clarify the origin of the phase transition near $T_p$, we carried out structural characterization by the SAED studies using TEM technique. Fig. 3 (a,c) and 3 (b,d) show, respectively, the SAED patterns with the electron incidence parallel to the [0001] and [10$\bar{1}$0] direction at 300 K and 100 K for EuCu$_{4-\delta}$P$_2$. It shows the same patterns above and below $T_p$. Fig. 3 (e) and 3 (f) show the SAED patterns of SrCu$_{4-\delta}$P$_2$ taken along zone axis [0001] at 300 K and 100 K, respectively. Except for the variations of the spot brightness intensity, no noticeable change of crystal structures in space group occurs, indicating that the transition due to any broken translational/rotational symmetry could be ruled out at $T_p$ in $Ln$Cu$_{4-\delta}$P$_2$.

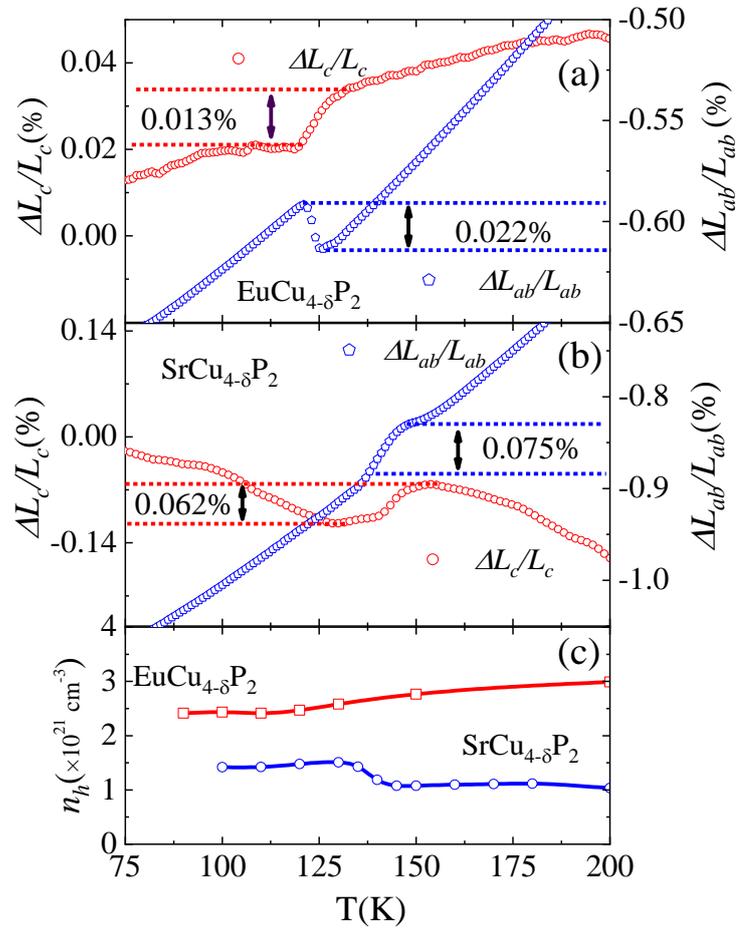

Figure 4. (a) and (b) Thermal expansion of lattice parameters along $c$-axis and in $ab$ plane in EuCu$_{4-\delta}$P$_2$ and SrCu$_{4-\delta}$P$_2$, respectively. (c) The carrier concentration of EuCu$_{4-\delta}$P$_2$ (red square line) and SrCu$_{4-\delta}$P$_2$ (blue dot line) extracted from Hall resistivity curves.

Our measurements mentioned above strongly demonstrate that the transition at $T_p$ in $Ln$Cu$_{4-\delta}$P$_2$ does not endure any type of magnetic or crystal structural transition, i.e., any mechanisms relative to broken spin ordering and crystal symmetry can be ruled out. Fig. 4(a) and (b) show the temperature dependences of thermal expansion ($\Delta L/L$) on the crystals both in the $ab$ plane and along $c$-axis of EuCu$_{4-\delta}$P$_2$ and SrCu$_{4-\delta}$P$_2$ respectively. We probed a slightly drop of $c$-axis about $\Delta L_c/L_c \sim 0.013\%$ while a simultaneous jump of $ab$ plane about $\Delta L_{ab}/L_{ab} \sim 0.027\%$ when the temperature decreases from 125 K down to 120 K in EuCu$_{4-\delta}$P$_2$. Similarly, a slightly drop of $c$-axis about $\Delta L_c/L_c \sim 0.062\%$ and $ab$ plane about $\Delta L_{ab}/L_{ab} \sim 0.075\%$ are observed in SrCu$_{4-\delta}$P$_2$ when the temperature decreases from 145 K down to 130 K in SrCu$_{4-\delta}$P$_2$. These subtle changes of the lattice parameters correspond to the $T_p$ transition at 120 K in EuCu$_{4-\delta}$P$_2$ and 140 K in SrCu$_{4-\delta}$P$_2$ determined from the temperature-dependent resistivity and the reciprocal susceptibility curves. Fig. 4(c) shows the temperature-dependent carrier concentration of EuCu$_{4-\delta}$P$_2$ (red square line) and SrCu$_{4-\delta}$P$_2$ (blue dot line) extracted from Hall resistivity measurements (Hall raw data of Hall resistivity is shown in Fig. S2 in the supplementary materials), it is seen that slight changes on the carrier concentration are detected at the corresponding transition temperatures of two samples, but no changes of the carrier type, indicating the electronic band structure has a little changes cross the $T_p$.

The λ-shaped hump anomaly of heat capacity near $T_p$ suggests that the $T_p$-transition belongs to a second order-like thermodynamic phase transition. Similar temperature-dependent resistivity, reciprocal susceptibilities and heat capacity have been also observed in EuAg$_4$As$_2$ [28]. The crucial difference here is that we did not observe noticeable variations at

SAED pattern across $T_p$ in our $Ln$Cu$_{4-\delta}$P$_2$, while a superlattice structure emerges below $T_p$ in EuAg$_4$As$_2$ by the x-ray diffraction precession study [27]. The conclusion of the second order transition also could be inferred by the result of the lattice parameter expanding 0.013% along $c$-axis and shrinking 0.022% in $ab$ plane in EuCu$_{4-\delta}$P$_2$, while it expands 0.062% and 0.075% in both directions in SrCu$_{4-\delta}$P$_2$ upon heating. However, this subtle lattice distortion is significantly smaller than a typical crystal structural transition with a lattice change on the order of ~0.5-1% [13,35].

Additionally, we have grown single crystals of EuCu$_{4-\delta}$P$_2$ with different growth processing, especially different Cu/P ratio in starting materials. The obtained EuCu$_{4-\delta}$P$_2$ single crystals show different $T_p$ varies from 120 K to 160 K (shown in Fig. S3-S5 in the supplementary materials), which means the $T_p$ is related to the Cu deficiency ratio δ. This strongly suggests that the $T_p$ transition should be driven by orbital from Cu $3d^9/3d^{10}$, which may account for the anomalies of $1/\chi$-$T$ ($\chi$-$T$) and carrier concentration in $Ln$Cu$_{4-\delta}$P$_2$. Hence, these iso-structural materials may provide a platform for research of orbital order and strongly correlated systems. To further figure out the underlying physics of the unique $T_p$ transition in $Ln$Cu$_{4-\delta}$P$_2$, resonance inelastic X-ray scattering (RIXS), angle-resolved photoemission spectroscopy (ARPES) or scanning tunneling spectrum (STS) measurements are helpful to reveal the coupling of charge, spin and orbit, electric band structure.

In conclusion, $Ln$Cu$_{4-\delta}$P$_2$ ($Ln$=Eu or Sr) endures a unique transition at $T_p$ ~120 K/140 K. It is a Landau type-II transition without crystal structure phase transition, charge order, and broken translational/rotational symmetry but only subtle lattice parameters change that expands 0.013% along $c$-axis while shrinks 0.022% in $ab$ plane with temperature decreasing in EuCu$_{4-\delta}$P$_2$ and expands only 0.062% along *c*-axis and 0.075% in *ab* plane in SrCu$_{4-\delta}$P$_2$. The transition at $T_p$ may originate from a subtle orbital ordering transition from Cu-P layer.

This work was supported by the National Key Research and Development Program of China (Grant No. 2021YFA1600201), the Natural Science Foundation of China (No. U19A2093, U2032214, U2032163, No. U1732274, No. 11904002, No. 11874359), and Collaborative Innovation Program of Hefei Science Center, CAS (Grant No. 2019HSC-CIP 001), and Youth Innovation Promotion Association of CAS (Grant No. 2021117), Natural Science Foundation of Anhui Province (No.1908085QA15). A portion of this work was supported by the High Magnetic Field Laboratory of Anhui Province.